\documentclass[aps,prl,reprint,superscriptaddress,amssymb,amsmath]{revtex4-1}
  \usepackage[utf8]{inputenc}
  \usepackage[T1]{fontenc}
  \usepackage[english]{babel}
  \usepackage{times}
  \usepackage{amssymb,amsmath,amsfonts,amsthm}
  \usepackage{mathtools}
  \usepackage{verbatim}
  \usepackage{enumerate}
  \usepackage{graphicx}
  \usepackage{hyperref}
  \usepackage{bbm}
  \usepackage{booktabs}
  
  \usepackage[caption=false]{subfig}
  \usepackage{mathrsfs}
  \usepackage{color}
  \usepackage{bm}
  \usepackage{natbib}
\usepackage{graphicx}
\usepackage{multirow}
\usepackage{bm}

\begin{document}
\title{ Experimental realization of sensitivity enhancement and suppression with exceptional surfaces}

\author{Guo-Qing Qin}
\thanks{These authors contributed equally to this work.}
\affiliation{State Key Laboratory of Low-Dimensional Quantum Physics and Department of Physics, Tsinghua University, Beijing 100084, P.R.China}

\author{Ran-Ran Xie}%
\thanks{These authors contributed equally to this work.}
\affiliation{State Key Laboratory of Low-Dimensional Quantum Physics and Department of Physics, Tsinghua University, Beijing 100084, P.R.China}
\author{Hao Zhang}%
\thanks{These authors contributed equally to this work.}
\affiliation{State Key Laboratory of Low-Dimensional Quantum Physics and Department of Physics, Tsinghua University, Beijing 100084, P.R.China}

\author{Yun-Qi Hu}%
\affiliation{State Key Laboratory of Low-Dimensional Quantum Physics and Department of Physics, Tsinghua University, Beijing 100084, P.R.China}
\author{Min Wang}%
\affiliation{State Key Laboratory of Low-Dimensional Quantum Physics and Department of Physics, Tsinghua University, Beijing 100084, P.R.China}
\author{Gui-Qin Li}%
\affiliation{State Key Laboratory of Low-Dimensional Quantum Physics and Department of Physics, Tsinghua University, Beijing 100084, P.R.China}

\author{Haitan Xu}%
\email{xuht@sustech.edu.cn}
\affiliation{Shenzhen Institute for Quantum Science and Engineering, Southern University of Science and Technology, Shenzhen, 518055, China}

\author{Fuchuan Lei}%
\email{fuchuan@chalmers.se}
\affiliation{Department  of  Microtechnology  and  Nanoscience, Chalmers  University  of  Technology, SE-41296  Gothenburg,  Sweden}

\author{Dong Ruan}%
\email{dongruan@mail.tsinghua.edu.cn}
\affiliation{State Key Laboratory of Low-Dimensional Quantum Physics and Department of Physics, Tsinghua University, Beijing 100084, P.R.China}

\author{Gui-Lu Long\textsuperscript{1,} }
\email{gllong@tsinghua.edu.cn}
\affiliation{Frontier Science Center for Quantum Information, Beijing, China}
\affiliation{Beijing National Research Center for Information Science and Technology, Beijing, China}
\affiliation{Beijing Academy of Quantum Information Sciences, Beijing, China}
\affiliation{School of Information, Tsinghua University, Beijing, China}
\date{\today}%
\begin{abstract}
By preparing a sensor system around isolated exceptional points, one can obtain a great enhancement of the sensitivity benefiting from the non-Hermiticity. However, this comes at the cost of reduction of the flexibility of the system, which is critical for practical applications. By generalizing the exceptional points to exceptional surfaces, it has been theoretically proposed recently that  enhanced sensitivity and flexibility can be combined. Here, we experimentally demonstrate an exceptional surface in a non-Hermitian photonic sensing system, which is composed of a whispering-gallery-mode microresonator and two nanofiber waveguides, resulting in a unidirectional coupling between two degenerate counter-propagating modes with an external optical isolator. The system is simple, robust, and can be easily operated around an exceptional surface. On the one hand, we observe sensitivity enhancement by monitoring the resonant frequency splitting caused by small perturbations. This demonstration of exceptional-surface-enhanced sensitivity paves the way for practical non-Hermitian sensing applications. On the other hand, we also show the suppression of frequency splitting around the exceptional surface for the first time.
\end{abstract}
\pacs{Valid PACS appear here}
\maketitle

\section{Introduction}
Exceptional points (EPs), known as spectral singularities in the non-Hermitian spectrum \cite{elganainy2018non-hermitian,ozdemir2019parity,feng2017non,miri2019exceptional} have been exploited recently for rich novel physics in both classical and quantum systems
 \cite{Bandreseaar4005,zhang2020synthetic,wang2020electromagnetically,zhong2020exceptional-point-based,xu2016topological,Jingwei2020,ruter2010observation,weimann2017topologically,lin2011unidirectional,zhang2016observation,chen2020revealing,kawabata2017information,lee2014entanglement,ACSEP,Wu878,0Observation,klauck,Quiroz-Juarez:19,naghiloo}. Unlike in Hermitian systems, the eigenstates at EPs coalesce. This unique eigenstate space reduction 
 implies that systems around EPs can be exploited as sensors by monitoring the splitting of the eigen energies or frequencies of the systems \cite{wiersig2014enhancing,wiersig2016sensors,Wiersig:20}, and this method has  been  demonstrated in experiments recently \cite{hodaei2017enhanced, chen2017exceptional,lai2019observation,0Non}. However, to develop this simple and elegant method for real sensing applications, it is usually challenging because that all parameters have to be precisely controlled for the non-Hermitian systems to approach EPs. Indeed, for example, two carefully etched fiber tips need to be operated simultaneously with nanometer precision in order to achieve an EP in a whispering gallery mode (WGM) resonator \cite{chen2017exceptional}. Because  these  EPs in parameter space are isolated, any unavoidable perturbation such as fabrication errors or environmental noise is likely to drive the sensor systems away from EPs, which becomes a major issue in practical applications of EP-based sensors.

\begin{figure}
\centering
\includegraphics[width=0.99\linewidth]{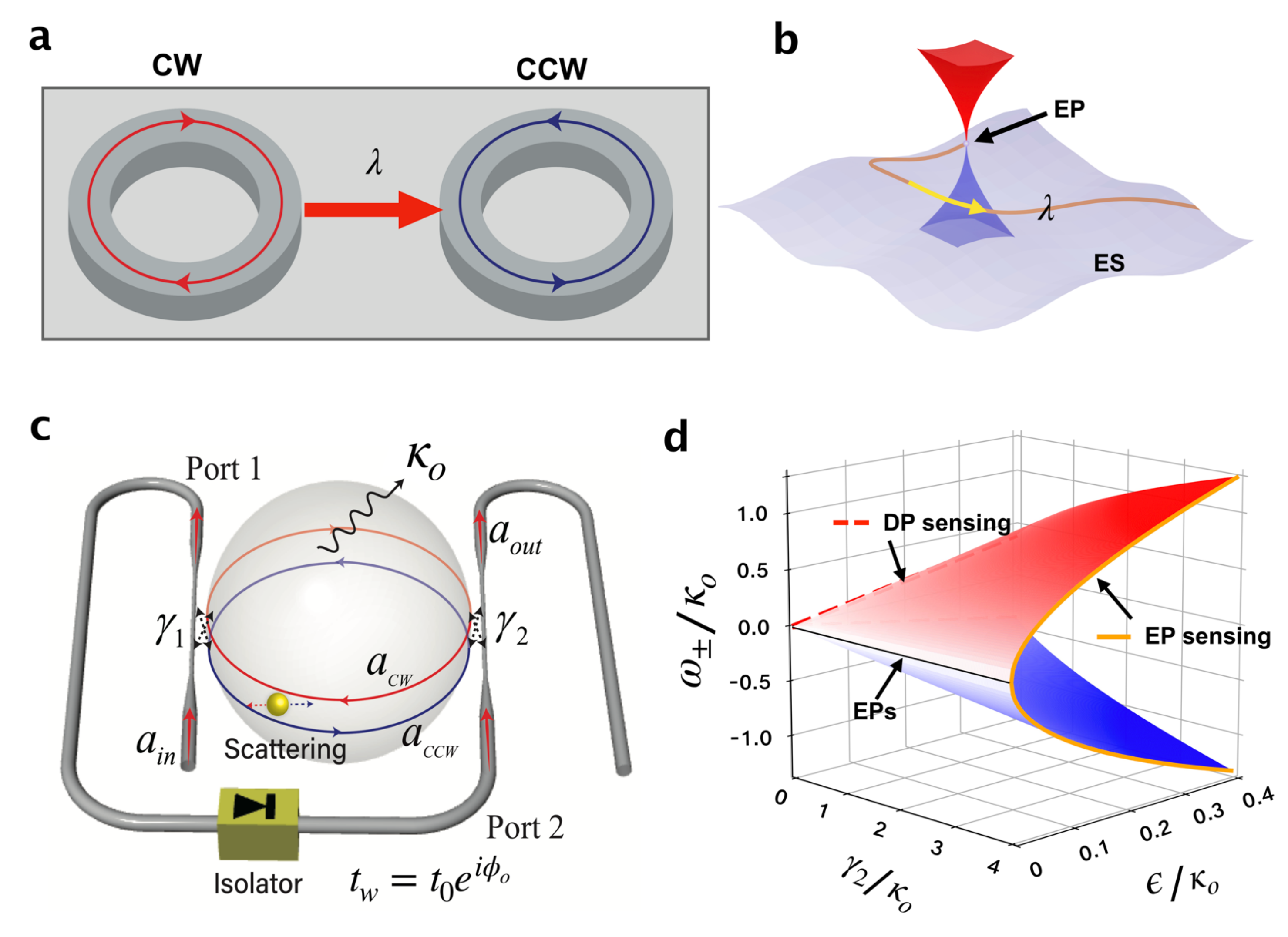}
\caption{($\mathbf{a}$) Schematic of configuring a non-Hermitian photonic system with a whispering gallery resonator where CW and CCW modes are unidirectionally coupled with coefficient $\lambda$. ($\mathbf{b}$) Conceptional illustration of an EP generalizing to a manifold in a higher dimensional physical parameter space, which is termed as the ES. ($\mathbf{c}$) The schematic of the experiment system. 
($\mathbf{d}$) The splitted eigenfrequencies on the basis of $\kappa_o$. The solid black line corresponds to a line of EPs with nonzero parameter $\gamma_2$. The splitting is linearly proportional to $\epsilon$ if $\gamma_2=0$, which is known as the diabolic-point (DP) sensing. While the system is around an EP for $\gamma_2>0$, the splitting scales as $\sqrt{\epsilon}$ represented by the orange line. In this plot,  $\phi_o=\frac{\pi}{2}$, $t_w =1$ and $\gamma_1=\gamma_2$.}
\label{fig1}
\end{figure}

To preserve the enhanced sensitivity while mitigating the difficulty of stabilizing the isolated EPs, a new type of non-Hermitian system which features an exceptional surface (ES) embedded in parameter space has been proposed recently \cite{zhong2019sensing}, and the ESs have just been observed on a magnon polariton platform composed of magnons and microwave photons \cite{zhang2019experimental}. Inspired by these remarkable works \cite{zhang2019experimental,Zhou19,okugawa2019topological,zhong2019sensing,budich2019symmetry-protected}, here we report an experimental implementation of a sensor based on ESs for the first time. A two-fold enhancment of sensitivity is observed. More importantly, we show that this enhancement can be achieved with flexible experimental parameters as the sensor operates around an ES, which paves the way for EP-based sensing in reality. Besides, we demonstrate the suppression of frequency splitting at ESs for the first time, which opens up new avenues for state control \cite{dong2012optomechanical,xu2016topological} in non-Hermitian systems.

\section{Results and Discussion}
\label{sec:Model}

\begin{figure*}[t]
\centering
\includegraphics[width=0.74\linewidth]{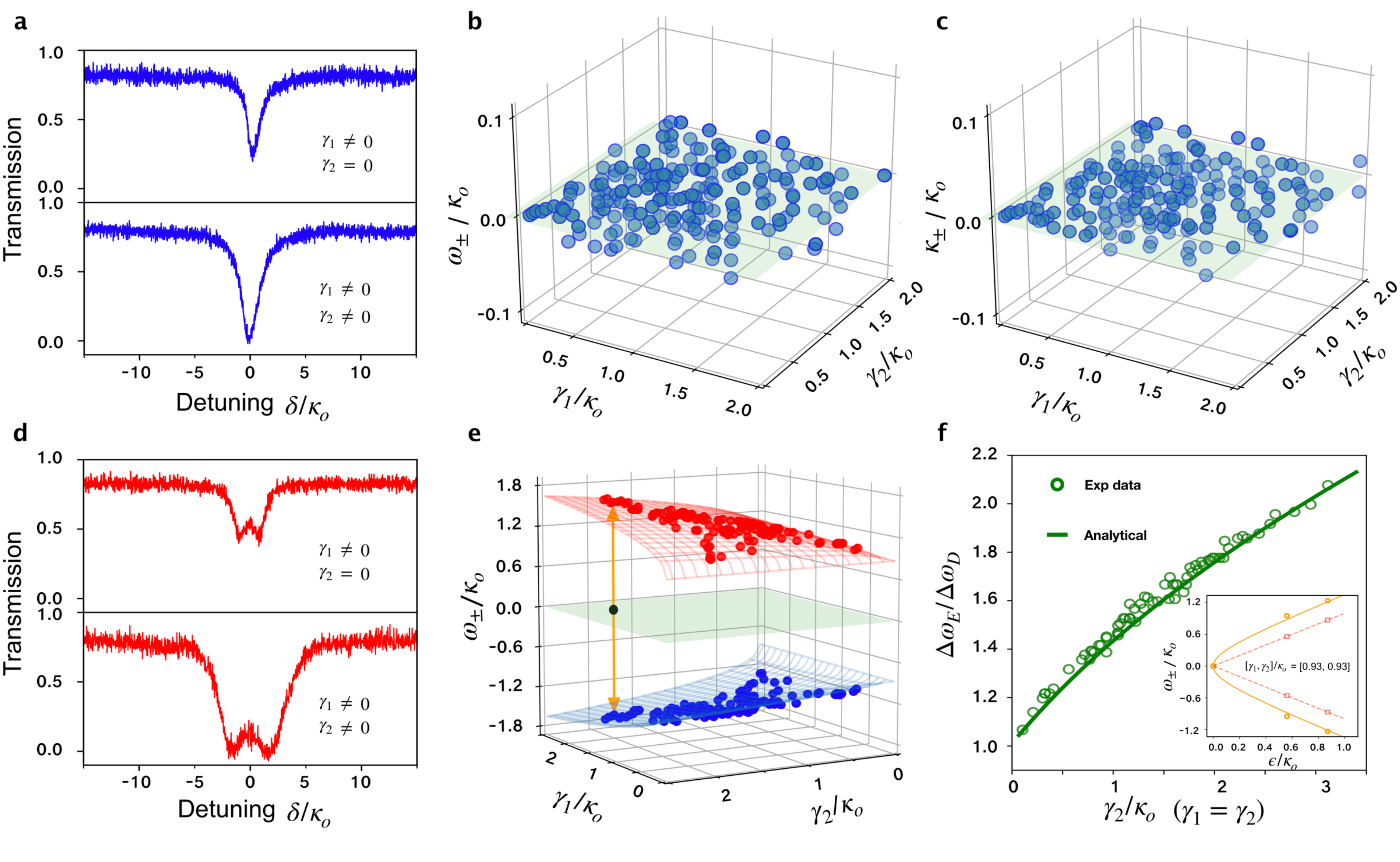}
\caption{Experimental observation of sensing at ES. ($\mathbf{a}$) The transmission spectra for a round microsphere ($\epsilon = 0$) coupled with a single taper or two tapers. The real parts ($\mathbf{b}$) and imaginary parts ($\mathbf{c}$) of the eigenfrequencies, showing an ES in the parameter space of [$\gamma_1$, $\gamma_2$]. The eigenvalues always coalesce when the coupling loss is changed. The blue dots are experimental data extracted from transmission spectra. ($\mathbf{d}$) The transmission spectra for a deformed microsphere ${\epsilon}/\kappa_o = 0.87$. The backscattering from CW mode induces mode splitting at DP. The mode splitting is enhanced when both tapered fibers couple with the cavity. ($\mathbf{e}$) Measured mode splitting based around an ES. Operating around different EPs on the ES could induce different sensitivity enhancement. ($\mathbf{f}$) Measured enhancement of mode splitting when the coupling strength ${\epsilon/\kappa_o} =0.87$. The green open circles are experimental data extracted from transmission spectra. The green solid line is analytic solution using experimental parameters. The inset present the behaviour of eigenfrequencies ${\omega}_{\pm}$ as a function of the coupling rate ${\epsilon}$. Measured results at the EP and DP are shown by open circles and squares, respectively. The solid curves are from analytical calculation using experimental parameters. Due to the loss between Port 1 and Port 2, $t_o$ is 0.895 in experiment.}
\label{fig2}
\end{figure*}

Our ES-enhanced sensor is built upon a high-quality-factor WGM resonator, which has been applied for a variety of sensing scenarios \cite{gavartin2012a,li2014single,Zhang:17,ward2018nanoparticle,yucavity}.  Conventionally, the sensing methods are based on the measurement of the frequency shift or linewidth broadening of high-Q resonances \cite{su2016label,ward2018nanoparticle,wan2018experimental}.  Different from Fabry–Pérot  resonators, the WGM resonators support pairs of clockwise (CW)- and counterclockwise (CCW)-propagating resonant modes with degenerate eigenfrequencies. This unique property provides an additional sensing mechanism, i.e., the  target information is imprinted in the signal of the induced resonant frequency splitting. This scheme has attracted increasing interest for its high sensitivity and self-referencing property \cite{zhu2010chip}. Moreover, it turns out the two-level system can be engineered to be at EPs, resulting in a further enhancement of the sensitivity \cite{chen2017exceptional}. However, as we mentioned above, it is never trivial to stabilize the system at an EP. Fortunately,  ESs can preserve the sensitivity enhancement in the presence of undesired perturbation. To operate the system at ESs, we introduce a unidirectional coupling between CW and CCW modes with coefficient $\lambda$, as shown in Figure. \ref{fig1}a. The asymmetric coupling between the two traveling modes with designed structures has been proposed in Ref. \cite{Bandreseaar4005,Ren18,zhong2019sensing}. We will show later that this unique coupling directly renders the system to be maintained at an EP without any additional requirements. As a result, a single EP generalizes to a continuous subspace in the physical parameter space, which is 
the so called ES conceptually illustrated in Figure. \ref{fig1}b. 

To achieve an ES experimentally, an optical tapered fiber is used to couple with a silica microsphere twice from two sides, see in Figure. \ref{fig1}c. In this configuration, the output from port 1 is reused as the input for port 2, while this coupling is not reciprocal because an optical isolator is inserted between the two ports. According to coupled-mode theory, the effective Hamiltonian $H_o$ can be written in the basis of two uncoupled modes $(a_{\rm cw}, a_{\rm ccw})^T$:
\begin{equation}
 H_o =\begin{pmatrix}  \omega_o -i \kappa & 0  \\ \lambda & \omega_o  -i \kappa \end{pmatrix} \label{eq2},
\end{equation}
where $\omega_o$ is the resonant frequency, and {{ $\kappa =(\kappa_o +\gamma_1+\gamma_2)/2$. $\kappa_o$ is the energy decay rate caused by the intrinsic loss; $\lambda=-i\sqrt{\gamma_1\gamma_2 }t_w$, which quantifies the coupling between CW and CCW cavity modes (see Sec.A of the Supplementary Information).}} $\gamma_1$ and $\gamma_2$ are the photon decay rates at the two coupling regions, respectively. Here $t_w=t_oe^{i\phi_o}$, which is a complex transmission coefficient between the coupling points with amplitude $t_o$ and phase accumulations $\phi_o$.

The target signal is a perturbation of the sensor, and without losing generality, it is given by an off-diagonal Hamiltonian 
\begin{equation}
 H_1 =\epsilon\begin{pmatrix}  0 & 1  \\ 1 & 0 \end{pmatrix} \label{2},
\end{equation}
where $\epsilon$ stands for the strength of perturbation. This Hamiltonian describes the interaction between CW and CCW modes in our system. For instance, the interaction could be caused by Rayleigh scattering, or deformation-induced coupling, etc.

To evaluate the sensitivity of the system, one can diagonalize the total Hamiltonian $H=H_o+H_1$ obtaining two complex eigenfrequencies 
\begin{align}
\chi\pm=\omega_o  -i\kappa \pm \sqrt{\epsilon(\epsilon+\lambda)}.\label{equ4}
\end{align} 
The real part $\omega_\pm$ and imaginary part $\kappa_\pm$ of the two complex eigenfrequencies on the basis of $\omega_o-i\kappa$ are given by 
\begin{align}
\omega_{\pm} &= \pm \bigg(\frac{\epsilon^2+\epsilon t_o \sqrt{\gamma_{1}\gamma_{2}}sin(\phi_o)}{2}\nonumber \\
&+\frac{\sqrt{\epsilon^4+2\epsilon^3 t_o \sqrt{\gamma_{1}\gamma_{2}}sin(\phi_o) +\epsilon^2 t^2_o \gamma_{1}\gamma_{2}}}{2} \bigg)^{\frac{1}{2}},\label{16}\\
\kappa_{\pm} &= \pm \bigg(\frac{-\epsilon^2-\epsilon t_o \sqrt{\gamma_{1}\gamma_{2}}sin(\phi_o)}{2}\nonumber \\
&+\frac{\sqrt{\epsilon^4+2\epsilon^3  t_o \sqrt{\gamma_{1}\gamma_{2}}sin(\phi_o) +\epsilon^2 t^2_o \gamma_{1}\gamma_{2}}}{2} \bigg)^{\frac{1}{2}}.\label{17}
\end{align}

If there is no unidirectional coupling, i.e., $\lambda=0$ (or $t_o=0$), this resonator has only a diabolic-point (DP) where the frequency splitting is $\Delta\omega\propto\epsilon$.

With a nonzero $\lambda$ and $\epsilon=0$, this Jordon-form 2$\times$2 Hamiltonian has just one linearly independent eigenvector
\begin{equation}
 \Phi_{EP} =\begin{pmatrix}  0 \\ 1  \end{pmatrix} \label{2},
\end{equation}
which means our sensor system is at an EP. In this case, one has $\Delta\omega\propto\sqrt{\epsilon}$ around the EP. Obviously, the sensitivity enhancement is giant for near zero $\epsilon$, which is shown clearly in Figure. \ref{fig1}d.

In the experiment, the eigenfrequencies are extracted from the transmission spectrum of the system  (see Sec. $\mathbf{B}$ of the Supplementary Information), obtained from the cavity input-output relation, i.e.,  $a_{out} = t_w ( a_{in}+\sqrt{\gamma_1} a_{\rm cw}) + \sqrt{\gamma_2} a_{\rm ccw}$. The silica microsphere resonator is used in our experiment with high optical quality factor modes \cite{2015Non}, $\kappa_o$ is 1.03 MHz when the silica resonator is excited with 1550.47 nm light. For an ideal microsphere resonator, there is no coupling between CW and CCW modes, i.e., $\epsilon=0$, the transmission spectra are composed of some Lorentzian dips \cite{cai2000observation,lei2020polarization}, and no mode splitting should be observed, see Figure. \ref{fig2}a. This is also true when the fiber is coupled with the resonator as the way shown in Figure. \ref{fig1}c. This holds the same for different coupling conditions, and as a result, an ES is formed in 
a parameter space 
with coordinates [$\gamma_1$, $\gamma_2$], as shown in  Figure. \ref{fig2}b and \ref{fig2}c. Certainly, the Rayleigh scattering always exists, but it might be too small to be measured due to the limited resonant linewidths.

Though an element $\lambda$ is added into the effective Hamiltonian due to the directional coupling (Equation. \ref{eq2}), the eigenfrequencies are still degenerate. To break the degeneracy for sensing, we introduce a small coupling between CW and CCW by modifying the resonator geometry with a $\rm CO_2$ laser \cite{xiao2007directional,oo2013evanescently}. It turns out that this permanent-modification method is quite helpful for studying ESs in experiments.

{{As is shown in Figure. \ref{fig2}d, the frequency splitting of the modified resonator can be clearly observed with a single taper coupling, i.e., $\gamma_2=0$. If the sphere is simultaneously coupled with the two tapers as we designed, an increased or decreased mode splitting can be observed for different $\phi_o$. 
To acquire the maximum mode splitting, i.e.,  the highest sensitivity, $\phi_o=\pi/2$ is required according to Equation. \ref{16}.
Figure. \ref{fig2}e shows the experimental data of the eigenfrequencies as a function of $\gamma_1$ and $\gamma_2$ with $\phi_o = \pi/2 \pm 0.04\pi$, where the Riemann surfaces are calculated using experiment parameters. It is found that the enhanced splitting can always be obtained for a large range of parameters $\gamma_1$ and $\gamma_2$,  this reflects the advantages of generalizing EPs to be ESs.}}

\begin{figure}
\centering
\includegraphics[width=0.96\linewidth]{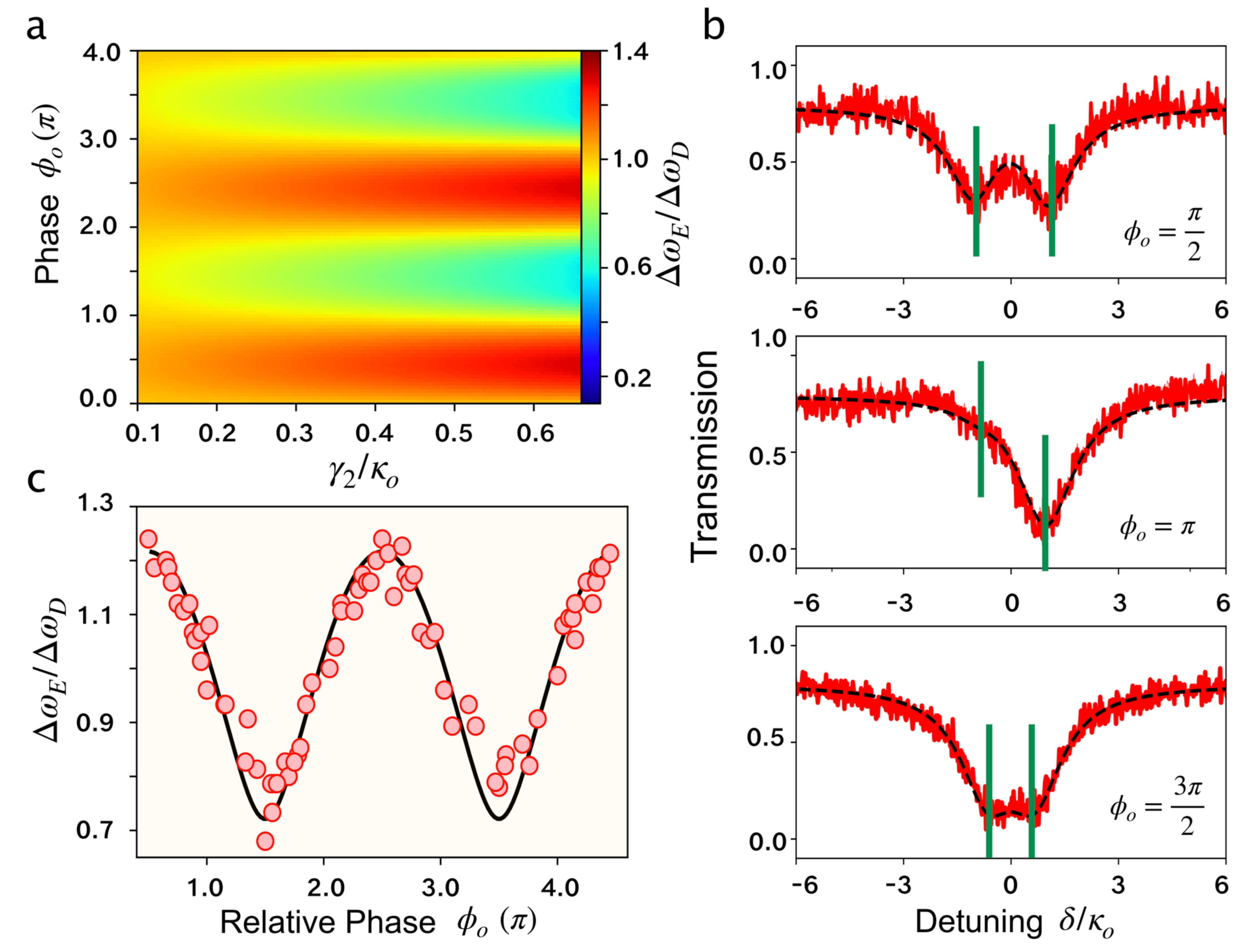}
\caption{Phase controlled frequency splitting. ($\mathbf{a}$) Simulation of frequency splitting as a function of phase $\phi_o$ and coupling loss $\gamma_2$. The coupling condition is maintained at $\gamma_1 = \gamma_2$. ($\mathbf{b}$) Transmission spectrum evolution by tuning the phase $\phi_o$ with ${\epsilon}/\kappa_o = 0.87$. ($\mathbf{c}$) Effect of phase $\phi_o$ on the frequency splitting with ${\epsilon}/\kappa_o = 0.87$. The solid black line is the analytical solution and the red dots denote experimental data extracted from transmission spectra. }
\label{fig3}
\end{figure}

Compared with the general non-Hermitian Hamiltonian, the Jordan-form Hamiltonian ensures that we can operate the system around EPs with variable parameters, i.e., ESs, which is advantageous for constructing a sensor in reality. In integrated photonic chips, the coupling coefficients are hard to be controlled precisely due to fabrication imperfection. In our system, the two coupling coefficients $\gamma_1$ and $\gamma_2$ are also susceptible to the mechanical 
perturbations and environment fluctuations.  Nevertheless, the system can be always operated at EPs as predicted. This system also enables us to access EP from DP continuously. Indeed, we show in Figure. \ref{fig2}f that the splitting increases with $\gamma_1$ ($\gamma_2$), and we observe an enhancement in frequency splitting by a maximum factor of 2. We also test with another deformed microsphere whose scattering coupling strength is different. As presented in the inset of Figure. \ref{fig2}f, the result is consistent with that in Figure. \ref{fig2}e. In principle, the ES-induced sensitivity enhancement is expected to be higher for sufficiently small perturbation. However, this is limited by the finite linewidths of resonance. This can be improved further by increasing quality factors.

\begin{figure*}[ht]
\centering
\includegraphics[width=0.8\linewidth]{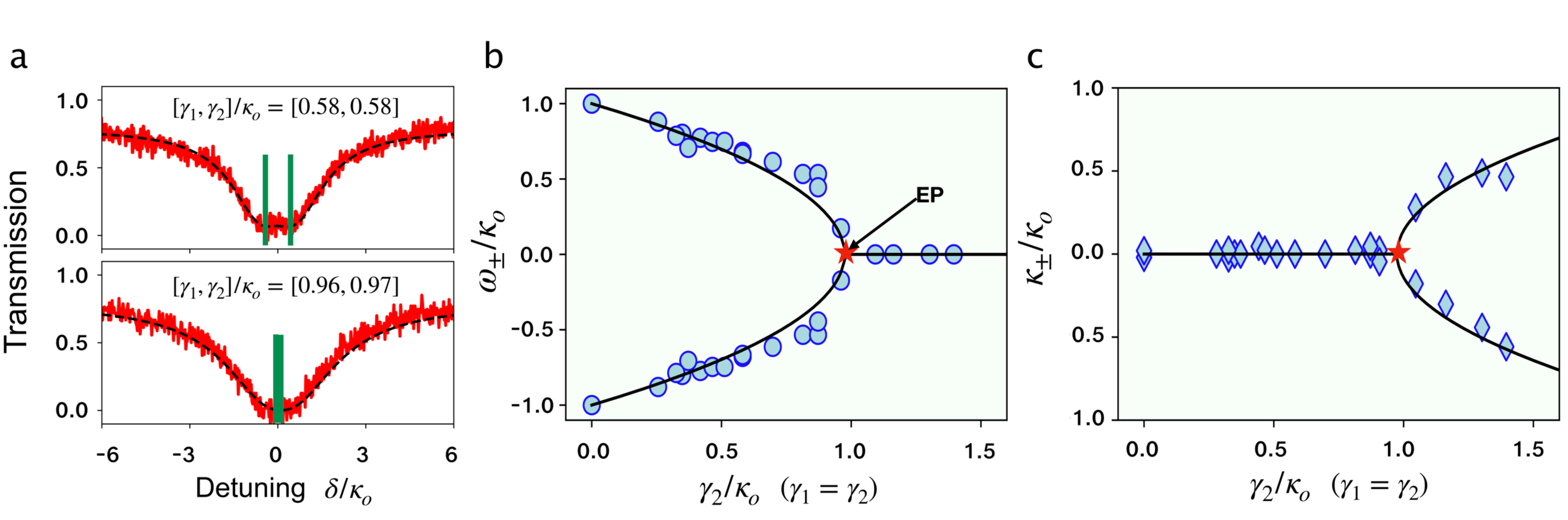}
\caption{Experimental observation of suppression of the mode splitting under the condition of $\phi_o = 3\pi/2$. Spectral evolution of power transmission for different coupling loss sets [$\gamma_1$, $\gamma_2$] are shown in ($\mathbf{a}$). The real parts ($\mathbf{b}$) and imaginary parts ($\mathbf{c}$) of the dynamical eigenfrequencies with ${\epsilon}/\kappa_o = 0.87$. Solid lines are calculated using the experimental coupling conditions extracted from transmission, and the circles and diamonds are experimental data. }
\label{fig4}
\end{figure*}

The ES-induced sensing can be intuitively understood as follows. 
In the case of a conventional DP sensing, the coupling between CW and CCW is solely mediated by the target \cite{zhu2010chip}, while in our case, it is the geometry deformation. In contrast, for an ES-sensing, there exists an additional coupling. As a result, the total coupling between two counter-propagating modes are the coherent superposition of two couplings \cite{wang2020electromagnetically}.

The above explanation implies that both enhancement and suppression can be achieved depending on their relative phase, which is demonstrated in Figure. \ref{fig3}. From Figure. \ref{fig3}a-c, one can see the frequency splitting varies as a function of $\phi_o$, as predicted by 
Equation. \ref{equ4}. This counter-intuitive phenomenon does not exist in regular isolated quantum and classical systems, where the eigenenergy and eigenfrequency splitting are irrelevant to the phase and solely determined by the strength of perturbation. The phase-controlled frequency splitting provides a new way of manipulation of non-Hermitian systems. Especially, with $\phi_o={3\pi}/{2}$, the frequency splitting can be significantly suppressed, as shown in Figure. \ref{fig4}a and \ref{fig4}b. For easier interpretation, we consider $\lambda+\epsilon =0$, the total Hamiltonian $H$ reduces to 
\begin{equation}
 H =\begin{pmatrix}  \omega_o -i \kappa & \epsilon  \\ 0 & \omega_o  -i \kappa \end{pmatrix} \label{eq10}.
\end{equation}
Therefore, the eigenvalues coalesce and an EP appears again in the Riemann surface, represented by the star in Figure. \ref{fig4}b.

\section{Conclusion} 
 In summary, we have experimentally realized an ES-based sensor in a WGM microcavity-waveguide coupled system. The realization of ES enables us to achieve sensitivity enhancement with a factor of two without the need of careful parameter optimization, which is benefical for constructing a practical sensor. The sensitivity enhancement are expected to be further improved with the assistance of optical gain \cite{hodaei2014parity,liu2018gain,li2014single}, especially for ultrasensitive applications. Thanks to the 
flexibility of this simple system, we have experimentally demonstrated  the frequency splitting can be continuously controlled by tuning the phase of the coupling coefficient. This unique property not only provides us new insight into non-Hermiation physical system but also a novel approach for manipulating the normal-mode resonance, such as Rayleigh scattering suppression \cite{Kim:19}, state conversion \cite{xu2016topological,yoon2018time-asymmetric}.
Last but not least, our system is also a good platform for the experimental exploration of higher-order EPs in photonic and other physical systems\cite{ding2016emergence}.

\noindent {\bf Acknowledgments.}
This work was supported by the National Natural Science Foundation of China under Grants (No.61727801); The National Key R$\&$D Program of China (2017YFA0303700); The Key R$\&$D Program of Guangdong province (2018B030325002); the National Natural Science Foundation of China under Grants (No.11974031); Tsinghua University Initiative Scientic Research Program; Beijing Advanced Innovation Center for Future Chip (ICFC).

\bibliographystyle{apsrev4-1}

\bibliography{sample}

\end{document}